\documentclass[aps,showpacs,twocolumn]{revtex4}

\usepackage{amssymb}
\usepackage{amsmath}

\input epsf

\begin{document}

\title{Production efficiency of Feshbach molecules in fermion systems}
\author{B. E. Dobrescu$^1$ and V. L. Pokrovsky$^{1,2}$}

\affiliation{$^1$Department of Physics, Texas A\&M University, College Station, Texas 77843-4242\\
$^2$Landau Institute of Theoretical Physics, Chernogolovka, Moscow region 142432, Russia}

\date{June 6, 2005}

\begin{abstract}
We present a consistent nonequilibrium theory for the production of
molecular dimers from a two-component quantum-degenerate atomic Fermi gas,
via a linear downward sweep of a magnetic field across a Feshbach
resonance. This problem raises interest because it is presently unclear as
to why deviations from the universal Landau-Zener formula for the transition
probability at two-level crossing are observed in the experimentally
measured production efficiencies. We show that the molecular conversion
efficiency is represented by a power series in terms of a dimensionless
parameter which, in the zero-temperature limit, depends solely on the
initial gas density and the Landau-Zener parameter. Our result reveals a
hindrance of the canonical Landau-Zener transition probability due to
many-body effects, and presents an explanation for the experimentally
observed deviations [K.E. Strecker \textit{et al.}, Phys. Rev. Lett.
{\bf 91}, 080406 (2003)].
\end{abstract}

\maketitle

\bigskip

The advances of last years in the experimental techniques of atomic and
molecular trapping and cooling, combined with the possibility of externally
tuning the inter-atomic interactions, have ushered in a series of novel
applications: the emergence of a molecular Bose-Einstein condensate (BEC)
from a Fermi gas \cite{JinBEC,Joachim}, observation of coherent oscillations
between an atomic condensate and molecules \cite{WiO-Donley}, formation and
propagation of matter-wave soliton trains \cite{HuletS}, and the examination
of Cooper pairing in the BCS-BEC crossover regime \cite{Regal2}.

Regardless of their role in further investigations, in all these experiments
diatomic molecules are being produced from an atomic BEC or a quantum
degenerate two-component Fermi gas either by sweeping a magnetic field
across a Feshbach resonance (FR) \cite{Bose,FermiR,FermiHulet,FermiCb,Hodby}
or by two-photon Raman photo-association \cite{Photo}. Therefore, the
understanding and control of the molecular production mechanism is of
special importance.

Notwithstanding the differences in the details of FR experiments, they all
show a growth of the molecular conversion efficiency (MCE) with the inverse
sweeping rate of the magnetic field, $\dot{B}^{-1}$, that saturates at values
less than 100$\%$ in the adiabatic regime.

The attempts aimed at explaining the dependence of MCE on $\dot{B}$,
resonance width, initial atomic density and temperature for two-component
Fermi systems can be broadly classified into two classes: i)semi-phenomenological scenarios 
\cite{LZscenarios,Ch-LZscenarios} that reduce the many-body physics to a two-atom
description modeled as a two-state Landau-Zener (LZ) system \cite{LZ}
corresponding, respectively, to the free two-atom scattering state and the
bound molecular state, and ii) numerical many-body calculations 
\cite{numerics1, numerics2} based on an effective Hamiltonian 
first proposed by E. Timmermans \textit{et al.} \cite{Timmermans}.


 Class i) is appealing by its use of simple and
intuitive physical pictures, but their predicted (temperature independent)
upper MCE limit of $50\%$ contradicts the
experimentally observed far greater values \cite{FermiCb,Hodby}. The
recent experimental work by E. Hodby \textit{et al.} \cite{Hodby}
also shows a pronounced $T$-dependence of this upper limit. 
The breakdown of the simple two-level LZ picture can be corrected only by
introducing suplimentary ad hoc assumptions in these semi-phenomenological scenarios,
whereas it emerges naturally from a bona fide many-body analysis (see below).
The work in Class ii) has shown, albeit under some simplifying
assumptions, the potential of Hamiltonian \cite{Timmermans} in
analyzing the temperature dependence of the MCE saturation in the
adiabatic regime.

In this Letter we study the atom-molecule conversion in
ultra-degenerate two-component Fermi gases subject to a
linear downward sweep of a magnetic field
across an $s$-wave FR, in the spirit of experiments
\cite{FermiR,FermiHulet,FermiCb,Hodby}. We focus on the
zero-temperature dynamics, and develop a nonequilibrium theory,
pertinent to both weak and strong atom-molecule coupling (measured
in Fermi energy units), which allows for a full account of the
effects of quantum statistics. The MCE is calculated in terms of
real-time Green functions (GF), and represented as a power series in
terms of a dimensionless parameter that depends only on the
initial gas density and the LZ parameter. An exact evaluation of
Feynman-Keldysh diagrams for second and fourth order processes
reveals a clear deviation from the LZ transition
probability at two-level crossing. This deviation, whose origins
reside solely in many-body effects, signals a suppression of the
LZ-predicted MCE even for moderately small values of
$\dot{B}^{-1}$, as observed experimentally in \cite{FermiHulet}.
Equally important, our MCE result doesn't display an a priori
upper limit of $50\%$ at $T=0$ as suggested in
\cite{Ch-LZscenarios}, though further work is necessary to
establish the correct limit and its $T$-dependence \cite{Hodby}.

The starting point of our analysis is the Hamiltonian \cite%
{Timmermans,Holland} describing a system of fermionic atoms FR-coupled to
bosonic molecules, $\hat{H}(t)=\hat{H}_{0}(t)+\hat{V}$, with
\begin{equation}
\hat{H}_{0}(t)=\sum_{\psi =a,b,f}\sum_{\vec{p}}\varepsilon _{\psi }(\vec{p}%
,t)\hat{\psi}^{\dagger }(\vec{p})\hat{\psi}(\vec{p}),  \label{eq1}
\end{equation}%
\begin{equation}
\hat{V}=\frac{g}{\sqrt{\mathcal{V}}}\sum_{\vec{p},\vec{q}}\left[ \hat{f}%
^{\dagger }\left( \vec{p}+\vec{q}\right) \hat{b}(\vec{q})\hat{a}(\vec{p})+%
\text{h.c.}\right] ,  \label{eq2}
\end{equation}%
\noindent where $\hat{a}^{\dagger }(\vec{p})$, $\hat{a}(\vec{p})$ and $\hat{b%
}^{\dagger }(\vec{p})$, $\hat{b}(\vec{p})$ are fermionic creation and
annihilation operators describing atoms of momentum $\vec{p}$ and
\textquotedblleft spins\textquotedblright\ $\uparrow $ $\left( a\right) $
and $\downarrow $ $\left( b\right) $, respectively, and $\hat{f}^{\dagger }(%
\vec{p})$, $\hat{f}(\vec{p})$ play the same role for the bosonic molecules.
Other quantities entering $\hat{H}$ are $\varepsilon _{\psi }(\vec{p},t)=%
\widetilde{\varepsilon }_{\psi }(\vec{p})-\mu _{\psi }B(t)$, with $\psi
=a,b,f$, where $\mu _{\psi }$ is the projection of the magnetic moment along
the direction of the magnetic field $B(t)$ with which interacts via
Zeeman coupling, and $\widetilde{\varepsilon }_{\psi }(\vec{p})$ is the
dispersion relation which accounts for the single-particle energy
renormalization due to nonresonant collisions, and simply reduces to the
kinetic energy $p^{2}/2m_{\psi }$ in a collisionless regime \cite{collision}%
; $g$ is the two-atom-molecule coupling \cite{coupling} which controls the
FR width and $\mathcal{V}$ is the volume of system. The free two-atom
scattering state and the molecular state (MS) have different spin
configurations and their coupling is mediated via the intra-atomic hyperfine
interaction \cite{Stoof-CC} which flips the electronic and nuclear spins of
one of the colliding atoms. Depending on the magnetically tuned energy
difference between the two states, the MS is quasi-bound (virtual) and
belongs to a closed scattering channel if its energy exceeds that of the
two-atom channel, becomes resonant with the latter when their energies are
equal, and turns truly bound when its energy is the lesser of the two.

In order to probe the MCE
dependence on $\dot{B}$ we evaluate real-time GF within the Keldysh
formalism (KF) \cite{Keldysh}. The method is based on the use of a closed
contour for time ordering, which runs from $-\infty $ to $+\infty $ and then
back to $-\infty $. Both branches of the contour propagate along the real
time axis and any point along them can be characterized by two parameters,
written compactly as $\tau ^{\gamma }$, with $\tau $ being the time variable
and $\gamma $ a bookkeeping index that distinguishes between the forward $%
\left( \gamma =+\right) $ and reverse $\left( \gamma =-\right) $ time
directions.
The basic quantities of KF are the contour-ordered real-time GF:%
\begin{equation}
i\mathcal{G}^{\alpha \beta }\left( \vec{p}_{1},\tau _{1};\vec{p}_{2},\tau
_{2}\right) =\left\langle \mathbf{T}_{c}\left[ \hat{\psi}_{H}\left( \vec{p}%
_{1},\tau _{1}^{\alpha }\right) \hat{\psi}_{H}^{\dagger }\left( \vec{p}%
_{2},\tau _{2}^{\beta }\right) \right] \right\rangle ,  \label{eq3}
\end{equation}%
\noindent with $\mathcal{G\equiv A}$, $\mathcal{B}$, $\mathcal{F}$ for $\psi
=a$, $b$, $f$, respectively, $\alpha ,$ $\beta =\pm $ and $\left\langle
\left( \cdots \right) \right\rangle \equiv $ Tr$\left[ \hat{\rho}%
(t_{0})\left( \cdots \right) \right] $; $\hat{\rho}(t_{0})$ is the initial
density operator at $t_{0}=-\infty $, $\hat{\psi}_{H}$ are the
Heisenberg-picture (HP) operators relative to $t_{0}$, and $\mathbf{T}_{c}$
is a contour-ordering operator. 
The corresponding free GF read $i%
\mathcal{G}_{0}^{\alpha \beta }\left( \vec{p}_{1},\tau _{1};\vec{p}_{2},\tau
_{2}\right) =\left\langle \mathbf{T}_{c}\left[ \hat{\psi}_{I}\left( \vec{p}%
_{1},\tau _{1}^{\alpha }\right) \hat{\psi}_{I}^{\dagger }\left( \vec{p}%
_{2},\tau _{2}^{\beta }\right) \right] \right\rangle $, where $\hat{\psi}%
_{I} $ are the interaction-picture (IP) operators relative to $t_{0}$.

In experiments \cite{FermiR,FermiHulet,FermiCb,Hodby} an ultracold
two-component Fermi gas is prepared as an incoherent mixture of equal
populations in each state, and extreme quantum-degeneracy, at temperatures
as low as $T\sim 0.05T_{F}$, has been reached \cite{Hodby}, where $T_{F}$ is
the Fermi temperature. In this regime the fall-off of the Fermi distribution
from $1$ to $0$ takes place in an extremely narrow energy interval $\sim
0.05\varepsilon _{F}$, where $\varepsilon _{F}$ is the Fermi energy, and the
fuzziness of the Fermi surface becomes virtually unimportant. In this vein
\cite{zeroT}, we take $\hat{\rho}(t_{0})=\left\vert \Phi _{0}\right\rangle
\left\langle \Phi _{0}\right\vert $, with $\left\vert \Phi _{0}\right\rangle
=\prod\limits^{<}\hat{a}^{\dagger }(\vec{p})\hat{b}^{\dagger }(\vec{p})|$VAC$%
\rangle $, where $|$VAC$\rangle $ is the vacuum state \cite{endisp}.
The average number of molecules at time $t$ is given by
\begin{equation}
\left\langle \hat{N}_{f}\right\rangle \left( t\right) =i\sum_{\vec{k}}%
\mathcal{F}^{+-}\left( \vec{k},t;\vec{k},t\right) ,  \label{eq4}
\end{equation}%
\noindent and, upon expressing the $\hat{\psi}_{H}$ operators in terms of
their IP form $\hat{\psi}_{I}$, $\hat{\psi}_{H}\left( t\right) =\hat{U}%
_{I}^{\dagger }(t,t_{0})\hat{\psi}_{I}\left( t\right) \hat{U}_{I}(t,t_{0})$,
and expanding the IP\ time-evolution operator $\hat{U}_{I}(t,t_{0})$ as a
formal series in the coupling constant $g$, the following systematic
expansion of $i\mathcal{F}^{+-}\left( \vec{k},t;\vec{k},t\right) $ ensues:
\begin{eqnarray}
&&i\mathcal{F}_{0}^{+-}\left( \vec{k},t;\vec{k},t\right) +\sum_{n=1}^{\infty
}\left( \frac{1}{i\hbar }\right) ^{n}\frac{1}{n!}  \notag \\
&&\times \sum_{\left\{ \gamma \right\} =\pm }\left( \gamma _{1}\cdots \gamma
_{n}\right) \int_{-\infty }^{+\infty }d\tau _{1}\cdots \int_{-\infty
}^{+\infty }d\tau _{n}  \label{eq5} \\
&&\times \left\langle \mathbf{T}_{c}\left[ \hat{V}_{I}\left( \tau
_{1}^{\gamma _{1}}\right) \cdots \hat{V}_{I}\left( \tau _{n}^{\gamma
_{n}}\right) \hat{f}_{I}\left( \vec{k},t^{+}\right) \hat{f}_{I}^{\dagger
}\left( \vec{k},t^{-}\right) \right] \right\rangle ,  \notag
\end{eqnarray}%
\noindent where $\hat{V}_{I}$ is the IP form of $\hat{V}$, and the sum $%
\sum_{\left\{ \gamma \right\} =\pm }$ runs over all $n$-tuples $\left(
\gamma _{1},\ldots ,\gamma _{n}\right) $ with $\gamma _{j}=\pm $.

Since $\hat{f}_{I}\left( \vec{p},t\right) \left\vert \Phi _{0}\right\rangle
=0$ for any $\vec{p}$ and $t$, and $\hat{V}_{I}\sim \hat{f}_{I}+\hat{f}%
_{I}^{\dag }$, it follows that only terms with even $n$ can have a
nonvanishing contribution in Eq. $\left( \ref{eq5}\right) $.

Due to the form of $\left\vert \Phi _{0}\right\rangle $, the free GF\ $%
\mathcal{G}_{0}^{\alpha \beta }\left( \vec{p}_{1},\tau _{1};\vec{p}_{2},\tau
_{2}\right) \varpropto \delta \left( \vec{p}_{1},\vec{p}_{2}\right) $, where
$\delta \left( \vec{p}_{1},\vec{p}_{2}\right) $ is the Kronecker delta, and
their expressions \cite{GFS} are
\begin{eqnarray}
i\mathcal{G}_{0}^{+-}\left( \vec{p};\tau _{1},\tau _{2}\right) &=&-\theta
\left( \varepsilon _{F}-\widetilde{\varepsilon }_{\psi }(\vec{p})\right) e^{%
\frac{i}{\hbar }\int_{\tau _{1}}^{\tau _{2}}\varepsilon _{\psi }(\vec{p}%
,\tau )d\tau },  \label{eq6-1} \\
i\mathcal{G}_{0}^{-+}\left( \vec{p};\tau _{1},\tau _{2}\right) &=&\theta
\left( \widetilde{\varepsilon }_{\psi }(\vec{p})-\varepsilon _{F}\right) e^{%
\frac{i}{\hbar }\int_{\tau _{1}}^{\tau _{2}}\varepsilon _{\psi }(\vec{p}%
,\tau )d\tau },  \label{eq6-2}
\end{eqnarray}
\noindent with $\mathcal{G\equiv A}$, $\mathcal{B}$ for $\psi =a$, $b$,
respectively, and
\begin{eqnarray}
i\mathcal{F}_{0}^{+-}\left( \vec{p};\tau _{1},\tau _{2}\right) &=&0,
\label{eq7-1} \\
i\mathcal{F}_{0}^{-+}\left( \vec{p};\tau _{1},\tau _{2}\right) &=&e^{\frac{i%
}{\hbar }\int_{\tau _{1}}^{\tau _{2}}\varepsilon _{f}(\vec{p},\tau )d\tau },
\label{eq7-2}
\end{eqnarray}
\noindent and finally $\mathcal{G}^{++}\left( \vec{p};\tau _{1},\tau
_{2}\right) =\theta \left( x\right) \mathcal{G}^{-+}\left( \vec{p};\tau
_{1},\tau _{2}\right) +\theta \left( -x\right) \mathcal{G}^{+-}\left( \vec{p}%
;\tau _{1},\tau _{2}\right) $, $\mathcal{G}^{--}\left( \vec{p};\tau
_{1},\tau _{2}\right) =\theta \left( x\right) \mathcal{G}^{+-}\left( \vec{p}%
;\tau _{1},\tau _{2}\right) +\theta \left( -x\right) \mathcal{G}^{-+}\left(
\vec{p};\tau _{1},\tau _{2}\right) $ for any $\mathcal{G\equiv A}$, $%
\mathcal{B}$, $\mathcal{F}$, where $x=\tau _{1}-\tau _{2}$ and $\theta
\left( x\right) $ is the Heaviside function.


\begin{figure}[tbp]
\begin{center}
\mbox{ \epsfxsize 3.2in\epsfbox{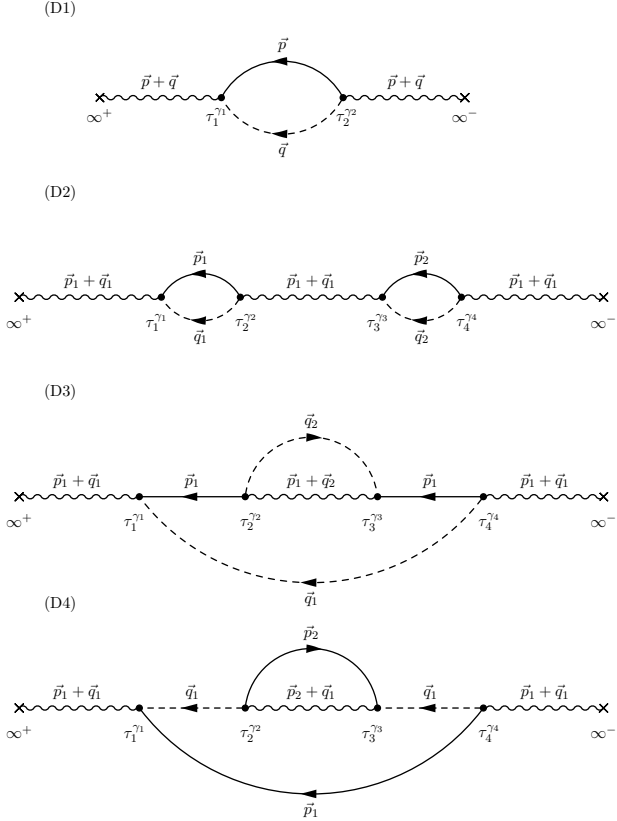}}
\end{center}
\caption{Feynman-Keldysh diagrams for second (D1) and fourth
order (D2 - D4) processes. The free Green functions are represented by
continuous lines for $a$-fermions ($\mathcal{A}_{0}^{\protect\gamma _{i}%
\protect\gamma _{j}}$), by dashed lines for $b$-fermions ($\mathcal{B}_{0}^{%
\protect\gamma _{i}\protect\gamma _{j}}$), and by wiggly lines for bosons ($%
\mathcal{F}_{0}^{\protect\gamma _{i}\protect\gamma _{j}}$).}
\label{diagr}
\end{figure}


Each average corresponding to the terms in Eq. $\left( \ref{eq5}%
\right) $ can be performed by means of a generalized version of Wick's
theorem in which the contractions are defined with respect to the
contour-ordering operator $\mathbf{T}_{c}$, and a diagram is associated 
with each way of contracting the field
operators into pairs \cite{Keldysh}. These diagrams have the same topology
as those occurring in the ordinary quantum field theory (OQFT) for systems
in equilibrium \cite{AGD}, the only difference being an additional label $%
\gamma =\pm $ that has to be attached to each interaction vertex.
As in OQFT, the disconnected diagrams corresponding to vacuum
polarization vanish \cite{Keldysh}, and only topologically
distinct diagrams need to be considered.

Since we are interested in analyzing the dependence of MCE on 
$\dot{B}^{-1}$, and not the behavior of the average number of molecules in time,
we set the initial
time of atomic gas preparation at $t_{0}=-\infty $, and the
molecule-counting time at $t_{m}=\infty $.
The diagrams representing the contribution from second and fourth order
processes to $\left\langle \hat{N}_{f}\right\rangle \left( \infty \right) $
are shown in Fig. $\left( \ref{diagr}\right) $. The (D1) diagrams contribute as
\begin{eqnarray}
&&\left( \frac{g}{\hbar \sqrt{\mathcal{V}}}\right) ^{2}\sum_{\vec{p},\vec{q}%
}\int_{-\infty }^{+\infty }d\tau _{1}\int_{-\infty }^{+\infty }d\tau _{2}
\notag \\
&&\times i\mathcal{F}_{0}^{-+}\left( \vec{p}+\vec{q};\infty ,\tau
_{1}\right) i\mathcal{F}_{0}^{-+}\left( \vec{p}+\vec{q};\tau _{2},\infty
\right)  \notag \\
&&\times i\mathcal{A}_{0}^{+-}\left( \vec{p};\tau _{1},\tau _{2}\right) i%
\mathcal{B}_{0}^{+-}\left( \vec{q};\tau _{1},\tau _{2}\right)  \notag \\
&=&\frac{N_{0}}{2}\times \left( 2\pi \frac{g^{2}}{\hbar ^{2}\dot{\Omega}}%
\frac{n_{0}}{2}\right) ,  \label{eq8}
\end{eqnarray}%
\noindent the contribution from the (D2) diagrams is
\begin{eqnarray}
&&-\left( \frac{g}{\hbar \sqrt{\mathcal{V}}}\right) ^{4}\sum_{\gamma
_{2},\gamma _{3}=\pm }\left( \gamma _{2}\gamma _{3}\right)  \notag \\
&&\times \sum_{\vec{p}_{1},\vec{q}_{1}}\sum_{\vec{p}_{2},\vec{q}_{2}}\delta
\left( \vec{p}_{1}+\vec{q}_{1},\vec{p}_{2}+\vec{q}_{2}\right)  \notag \\
&&\times \int_{-\infty }^{+\infty }d\tau _{1}\cdots \int_{-\infty }^{+\infty
}d\tau _{4}\text{ }i\mathcal{F}_{0}^{++}\left( \vec{p}_{1}+\vec{q}%
_{1};\infty ,\tau _{1}\right)  \notag \\
&&\times i\mathcal{F}_{0}^{\gamma _{2}\gamma _{3}}\left( \vec{p}_{1}+\vec{q}%
_{1};\tau _{2},\tau _{3}\right) i\mathcal{F}_{0}^{--}\left( \vec{p}_{1}+\vec{%
q}_{1};\tau _{4},\infty \right)  \notag \\
&&\times i\mathcal{A}_{0}^{+\gamma _{2}}\left( \vec{p}_{1};\tau _{1},\tau
_{2}\right) i\mathcal{B}_{0}^{+\gamma _{2}}\left( \vec{q}_{1};\tau _{1},\tau
_{2}\right)  \notag \\
&&\times i\mathcal{A}_{0}^{\gamma _{3}-}\left( \vec{p}_{2};\tau _{3},\tau
_{4}\right) i\mathcal{B}_{0}^{\gamma _{3}-}\left( \vec{q}_{2};\tau _{3},\tau
_{4}\right)  \notag \\
&=&\frac{N_{0}}{2}\times \frac{17}{105}\left( 2\pi \frac{g^{2}}{\hbar ^{2}%
\dot{\Omega}}\frac{n_{0}}{2}\right) ^{2},  \label{eq9}
\end{eqnarray}%
\noindent and each of the diagrams (D3) and (D4) contributes equally as
\begin{eqnarray}
&&\left( \frac{g}{\hbar \sqrt{\mathcal{V}}}\right) ^{4}\sum_{\gamma
_{2},\gamma _{3}=\pm }\left( \gamma _{2}\gamma _{3}\right) \sum_{\vec{p}_{1},%
\vec{q}_{1}}\sum_{\vec{p}_{2}}  \notag \\
&&\times \int_{-\infty }^{+\infty }d\tau _{1}\cdots \int_{-\infty }^{+\infty
}d\tau _{4}\text{ }i\mathcal{F}_{0}^{++}\left( \vec{p}_{1}+\vec{q}%
_{1};\infty ,\tau _{1}\right)  \notag \\
&&\times i\mathcal{F}_{0}^{\gamma _{2}\gamma _{3}}\left( \vec{p}_{2}+\vec{q}%
_{1};\tau _{2},\tau _{3}\right) i\mathcal{F}_{0}^{--}\left( \vec{p}_{1}+\vec{%
q}_{1};\tau _{4},\infty \right)  \notag \\
&&\times i\mathcal{B}_{0}^{+\gamma _{2}}\left( \vec{q}_{1};\tau _{1},\tau
_{2}\right) i\mathcal{B}_{0}^{\gamma _{3}-}\left( \vec{q}_{1};\tau _{3},\tau
_{4}\right)  \notag \\
&&\times i\mathcal{A}_{0}^{+-}\left( \vec{p}_{1};\tau _{1},\tau _{4}\right) i%
\mathcal{A}_{0}^{\gamma _{3}\gamma _{2}}\left( \vec{p}_{2};\tau _{3},\tau
_{2}\right)  \notag \\
&=&-\frac{N_{0}}{2}\times \frac{1}{2}\left( 2\pi \frac{g^{2}}{\hbar ^{2}\dot{%
\Omega}}\frac{n_{0}}{2}\right) ^{2},  \label{eq10}
\end{eqnarray}%
\noindent where $\hbar \dot{\Omega}\equiv (\mu _{f}-\mu _{a}-\mu _{b})\dot{B}
$, $N_{0}$ is the total number of atoms present in the system before the
magnetic field is applied, and $n_{0}=N_{0}/\mathcal{V}$ is the initial
density.

The evaluation of all integrals entering Eqs. $\left( \ref{eq8}\right)
-\left( \ref{eq10}\right) $ can be carried out exactly, without any
supplementary assumptions, and a presentation of the steps involved is
beyond the scope of this letter.

Upon collecting results and introducing the notation $\Gamma \equiv 2\pi \xi
_{LZ}\left( \mathcal{V}\frac{n_{0}}{2}\right) $, where $\xi _{LZ}=\frac{g^{2}%
}{\mathcal{V}\hbar ^{2}\dot{\Omega}}$ is the canonical LZ\ parameter \cite%
{LZ}, we obtain
\begin{equation}
\text{MCE}=\frac{2\left\langle \hat{N}_{f}\right\rangle \left( \infty
\right) }{N_{0}}=\Gamma -\frac{88}{105}\Gamma ^{2}+\mathcal{O}\left( \Gamma
^{3}\right) .  \label{eq11}
\end{equation}

The $n$-th term of this series is represented by the set of Feynman-Keldysh
diagrams containing $2n$ vertices. A similar approach, i.e. the development
of a formal solution in terms of an infinite series in powers of the
coupling constant, with further calculation of every term and summation of
the resulting algebraic series, has been employed to exactly calculate the
probability of nonadiabatic transitions in a multiple-crossing LZ model \cite%
{Kayanuma}, and it constitutes a powerful alternative to
Landau's method \cite{LZ} of analytic continuation 
in complex time.

Eq. $\left( \ref{eq11}\right) $ reveals deviations from the
universal two-level LZ formula \cite{LZ}, and also from 
the phenomenological correction 
proposed in \cite{Ch-LZscenarios} as
\begin{equation}
\eta \left( 1-e^{-\Gamma }\right) =\eta \left( \Gamma -\frac{1}{2}\Gamma
^{2}+\mathcal{O}\left( \Gamma ^{3}\right) \right) ,  \label{eq12}
\end{equation}
\noindent where $\eta \leq 50\%$ is a constant that depends on the initial
populations of the two-component Fermi gas.

Since $\frac{88}{105}>\frac{1}{2}$, Eq. $\left( \ref%
{eq11}\right) $ shows that, as $\dot{B}^{-1}$ increases, the MCE
grows slower then predicted by the LZ formula, and this behavior
is experimentally supported \cite{FermiHulet}. 
In our theory,
the approach towards saturation is not due to a 
mere contraction of the LZ formula by a multiplicative factor determined 
solely by the initial state preparation, as proposed in
the LZ scenarios \cite{LZscenarios,Ch-LZscenarios}, but has a rather dynamical nature as 
the atom-molecule conversion takes place in a many-body medium in which the 
effects of quantum statistics play a crucial role.

Examination of
higher order diagrams indicates that MCE is a function depending
solely on the parameter $\Gamma $.
Therefore, in the extreme adiabatic regime, corresponding to
$\Gamma \rightarrow \infty $, MCE must have a universal limit at
$T=0$ which, unlike in the phenomenological result $\left( \ref{eq12}\right) $, 
is not a priori bounded by $50\%$. In practice, as the experiments are carried 
out at finite $T$ and $\Gamma $, the smearing of the Fermi
surface when $T$ approaches $T_{F}$, and
the quantum degeneracy reaches its lower limit, must be taken into account \cite{zeroT} for
analyzing the $T$-dependence of the MCE saturation
\cite{Hodby}.

In conclusion, we have considered the MCE for a hyperfine-induced
$s$-wave FR 
and developed a consistent many-body nonequilibrium theory, 
based on the real-time GF approach, in which all atomic and
molecular states are included, and the effects of quantum
statistics are fully accounted for. 

We demonstrated, by analytically evaluating
the MCE up to fourth order in the hyperfine coupling constant,
that the canonical LZ formula at two-level crossing is violated in
this system due to many-body effects which systematically decrease
the LZ transition probability, even for moderately small values of
$\dot{B}^{-1}$. 
This result indicates that
in degenerate Fermi gases the effects of quantum statistics near a FR 
play a crucial role, and the singling out of independent two-atom pairs
from an ensemble of delocalized indistinguishable particles, as proposed in
the LZ scenarios \cite{LZscenarios,Ch-LZscenarios}, is untenable.

We acknowledge support by the DOE Grant No. DE-FG03-96ER45598, and by NSF
under the Grant No. DMR-0321572.


\begin{references}
\bibitem{collision} In our terminology, the term collision refers to scattering processes
that cannot alter the number of atoms/molecules.
\bibitem{coupling} $g\sim \epsilon _{hf}a^{3/2}$, where $\epsilon _{hf}$ is the
strength of the hyperfine interaction responsible for the coupling bewteen
the electronic and nuclear spins, and $a$ is the characteristic size of the
molecule.
\bibitem{zeroT} We leave the extension to $T>0$ for future work.
\bibitem{endisp}
$\prod\limits^{<}$
represents a product over momenta $\vec{p}$, subject to
restriction $0\leq \widetilde{\varepsilon }(\vec{p})\leq \varepsilon _{F}$.
Since the two equally populated components of the Fermi gas correspond to two different internal states
of the same atom species, we take the same dispersion for them, i.e.
$\widetilde{\varepsilon }_{a}(\vec{p})=\widetilde{\varepsilon }_{b}(\vec{p}%
)\equiv \widetilde{\varepsilon }(\vec{p})$.
\bibitem{GFS} Since $\mathcal{G}_{0}^{\alpha \beta }\left( \vec{p}_{1},\tau
_{1};\vec{p}_{2},\tau _{2}\right) \varpropto \delta \left( \vec{p}_{1},\vec{p%
}_{2}\right) $, in order to simplify notation we write $\mathcal{G}%
_{0}^{\alpha \beta }\left( \vec{p}_{1};\tau _{1},\tau _{2}\right) $ instead.
\end{references}
\end{document}